\documentclass{vgtc}                         % final (conference style)
\havecopyrightspacefalse

\usepackage{mathptmx}
\usepackage{graphicx}
\usepackage{times}

\usepackage[utf8]{inputenc}
\usepackage[T1]{fontenc}

\onlineid{0}

\vgtccategory{Research}

\vgtcinsertpkg

\title{A Tangible Volume for Portable 3D~Interaction}

\author{Paul Issartel\thanks{e-mail: \{paul.issartel,mehdi.ammi\}@limsi.fr}\\ %
        \parbox{1.4in}{\scriptsize \centering LIMSI-CNRS \\ Univ. Paris-Sud} %
\and Lonni Besançon\thanks{e-mail: \{lonni.besancon,tobias.isenberg\}@inria.fr}\\ %
     \parbox{1.4in}{\scriptsize \centering INRIA Saclay\\ Univ. Paris-Sud} %
\and Tobias Isenberg\footnotemark[2]\\ %
     \parbox{1.4in}{\scriptsize \centering INRIA Saclay} %
\and Mehdi Ammi\footnotemark[1]\\ %
     \parbox{1.4in}{\scriptsize \centering LIMSI-CNRS \\ Univ. Paris-Sud}}

\teaser{
 \centering
 \includegraphics[width=\linewidth]{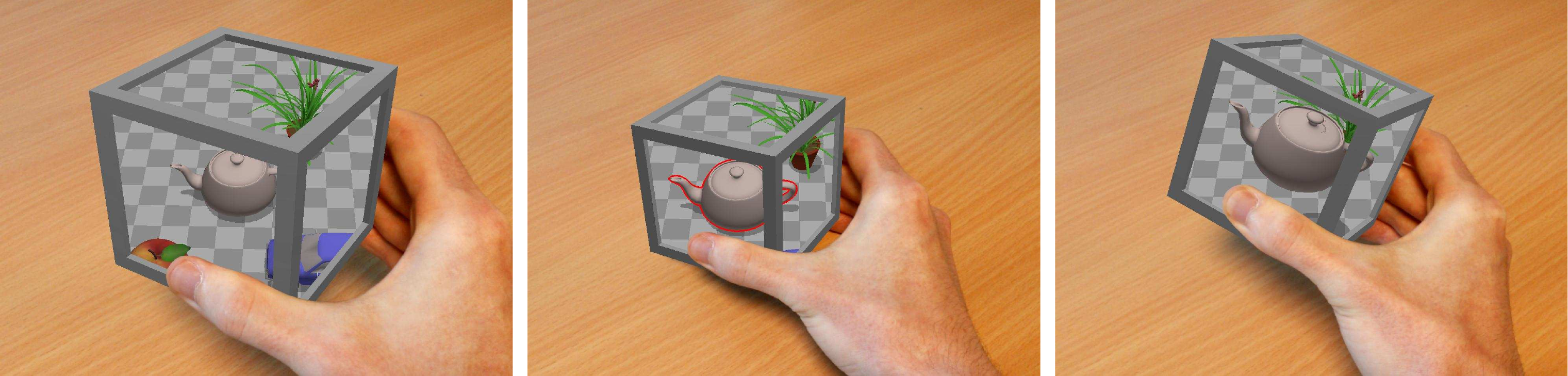}
 \caption{%
    Our concept of a \emph{tangible volume} consists of a fully portable and self-contained device, entirely covered with screens. A virtual scene can be seen ``through'' the volume of the device~(left image). This volume can be directly positioned within the virtual scene~(middle image), and used to grasp and manipulate virtual objects~(right image).
    A cubic-shaped device is shown here for illustration, but other volume shapes can be used.%
  }
 \label{fig:teaser}
}

\abstract{%
We present a new approach to achieve tangible object manipulation with a single, fully portable and self-contained device. Our solution is based on the concept of a ``tangible volume''. We turn a tangible object into a handheld fish-tank display. The tangible volume represents a volume of space that can be freely manipulated within a virtual scene. This volume can be positioned onto virtual objects to directly grasp them, and to manipulate them in 3D~space. We investigate this concept through two user studies. The first study evaluates the intuitiveness of using a tangible volume for grasping and manipulating virtual objects. The second study evaluates the effects of the limited field of view on spatial awareness. Finally, we present a generalization of this concept to other forms of interaction through the surface of the volume.%
}


\CCScatlist{
  \CCScat{H.5.1}%
  {Information Interfaces And Presentation}%
  {Multimedia Information Systems}%
  {Artificial, augmented, and virtual realities}
}

\usepackage[protrusion,expansion,tracking,kerning,spacing]{microtype}
\microtypecontext{spacing=nonfrench}
\SetTracking{encoding={*}, shape=sc}{0}
\usepackage{amsmath}
\usepackage{textcomp}
\usepackage{color}
\usepackage[normalem]{ulem}
\usepackage{enumitem}
\usepackage{url}
\usepackage{flushend}

\begin{document}


\firstsection{Introduction}

\maketitle

%% \section{Introduction}

Interaction with 3D~data, and in particular object manipulation~\cite{bowman04} (selecting, translating and rotating 3D~virtual objects) is of major importance in many fields, such as scientific visualization, prototyping and gaming. Until recently, these tasks were often carried out by advanced users on fixed workstations. However, with the development of mobile computing, people are now expecting to perform these tasks anywhere, with minimal set-up and learning. Thus, there is a need for an interface that is not only \emph{natural and efficient} for 3D~manipulation, but also truly \emph{portable and self-contained}.

Tangible User Interfaces~(TUIs) represent a promising approach for 3D~manipulation. They consist of physical objects, or tangible objects, that serve as real-world representations of digital data. In many~TUIs, tangible objects are used as physical ``handles'' for virtual objects. These interfaces take advantage of the user's skills in interacting with physical objects~\cite{ishii08}, making them an attractive solution to support manipulation tasks in a natural and efficient way.

However, interacting with virtual objects also requires visual feedback. Tangible~UIs for 3D~manipulation often rely on an external and fixed monitor for visual output~\cite{hinckley94}. Hence, such interfaces cannot be considered portable. Others use mobile devices as a portable display surface~\cite{issartel14-2}. Although these interfaces are portable, they still consist of multiple separate pieces, which always have to be handled and carried together. They are not self-contained. Some~TUIs take a further step and use the mobile display as a physical handle~\cite{henrysson05}. The device itself replaces the external tangible objects, thereby reducing the entire interface to a single portable and self-contained object. This solution, however, does not eliminate the distance between the display and the virtual objects. Since the mobile display now serves as the tangible handle, this separation creates problems during manipulation, such as a shifted center of rotation or manipulated objects leaving the field of view.

In this paper, we present a different approach. Rather than turning a mobile device into a tangible handle, we turn a tangible object into a display. We cover the surface of a tangible object with multiple screens, and use fish-tank rendering to display part of the virtual scene ``through'' the object. The object then becomes a tangible representation of a \emph{volume} of virtual space. We introduce a ``grasping'' metaphor that consists in positioning this volume directly onto a virtual object, and pressing the fingers to pick up the object with the tangible volume. Thus, the separation between the tangible handle and the manipulated virtual object disappears. By solving the issues caused by this separation, our solution makes it possible to preserve the advantages of tangible manipulation in a fully portable and self-contained device. Our contributions in this work are therefore:
\begin{itemize}[noitemsep,topsep=1mm]
\item a new concept of a portable and self-contained device that allows users to directly interact with a volume of virtual space,\\[-2.7mm]
\item a grasping metaphor that demonstrates an application of this concept to 3D~manipulation,\\[-2.7mm]
\item a partial prototype and two experimental studies to investigate several usability aspects of this mode of interaction, and\\[-2.7mm]
\item a generalization of our concept to other forms of interaction through the surface of the volume.
\end{itemize}

\section{Background}

\subsection{Tangible User Interfaces for 3D manipulation}

The main idea behind Tangible User Interfaces~(TUIs) is to give physical form to digital data~\cite{ishii08,ishii97}. This is accomplished through the use of real-world objects---called \emph{tangible objects}---that represent the digital information. One particular type of~TUI are Graspable User Interfaces~\cite{fitzmaurice96}, in which tangible objects serve as physical ``handles'' for virtual objects. Each handle can be attached to a virtual object. Users can then manipulate a virtual object by directly manipulating the corresponding tangible object. By taking advantage of the user's preexisting skills in manipulating real-world objects~\mbox{\cite{fitzmaurice96,ishii08}}, this interaction mode provides a natural and immediately efficient approach to 3D~manipulation.

A number of~TUIs have been designed around this concept. Hinckley~et~al.~\cite{hinckley94} used passive tangible objects (``props'') tracked in mid-air to position and orient 3D~medical data, The result was displayed on a separate computer screen. Similar interfaces were later proposed by Qi~and~Martens~\cite{qi05} and Jackson~et~al.~\cite{jackson13} with wireless tracking. The Personal Space Station~\cite{mulder02} is a situated virtual reality system that consists of a projector and a reflective screen. Users interact with the virtual 3D~objects by manipulating tangible objects behind the screen. A limitation of these interfaces, however, is that they all require an external monitor or projector. This makes them essentially fixed installations. The lack of portability creates many constraints. Users need to go to a dedicated place in order to use the system. It cannot be carried between offices or brought home. Nowadays, with the rise of mobile computing, such constraints appear even more limiting. There is a need for a truly portable interface that would offer the same advantages as fixed~TUIs. Thus, our first requirement is that the entire interface should be \emph{portable}.

These constraints can be addressed by making the display portable. Since mobile devices---such as smartphones and tactile tablets---have become widespread, they represent a readily-available solution for use as a portable display surface. Some researchers have combined mobile device with tangible objects~\cite{avrahami11,issartel14-2,liang13}, resulting in portable~TUIs. However, these interfaces also consist of multiple independent pieces: the mobile device and the tangible objects, all of which need to be carried together at all times. Whenever one of the tangible objects is missing, or when the mobile device is separated from the tangible objects, the interface becomes less functional or even unusable. The portability advantage is reduced due to the inconvenience of having to carry and keep the multiple pieces together. We can thus identify another requirement: the interface should not only be portable but should also consist of a single, \emph{self-contained}~device.

\subsection{Mobile device as tangible handle}

In addition to its role as a portable display surface, the mobile device itself can be seen as a tangible object. Several researchers have proposed to use the mobile device as tangible handle~\cite{henrysson05,marzo14,mossel13} to manipulate virtual objects in the space behind the device (\mbox{generally} in augmented reality). This eliminates the need for external tangible props, resulting in an interface that is fully portable and self-contained in a single object: the mobile device.

\begin{figure}[t]
  \centering
  \includegraphics[width=.95\linewidth]{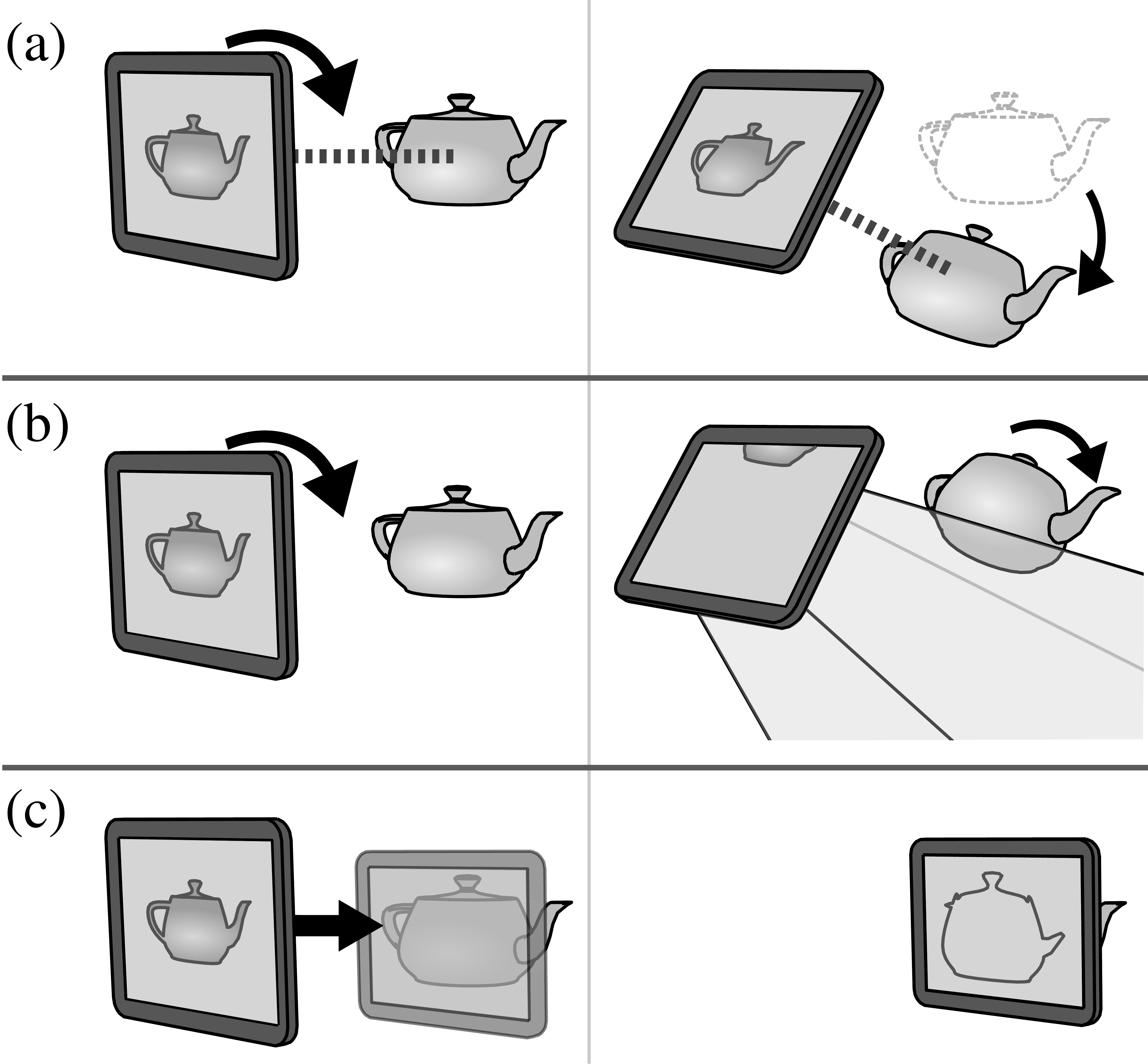}
  \caption{Manipulation issues that arise when a typical mobile device is used as a tangible handle:\enskip%
    (a)\;the virtual object does not rotate about its own center;\enskip%
    (b)\;the virtual object leaves the field of view;\enskip%
    (c)\;when trying to reduce the distance to avoid the above issues, the virtual object is clipped.}
  \label{fig:mobile-device-issues}
\end{figure}

Although this solution fully meets the portability requirements, it is not, however, without drawbacks. Unlike previous approaches, there is now a separation between the virtual objects (behind the device) and the tangible handle (the device itself). This separation leads to problems during manipulation. In the techniques proposed by Henrysson~et~al.~\cite{henrysson05} and Marzo~et~al.~\cite{marzo14}, the manipulated object is fixed relative to the device. Since the center of rotation is now located on the device, it becomes difficult to rotate a virtual object without also translating it~(Figure~\ref{fig:mobile-device-issues}(a)). In the HOMER-S technique~\cite{mossel13}, translations and rotations are separately applied to the manipulated object, which makes it easier to rotate the object about itself. On the other hand, the object is no longer fixed relative to the mobile device and can thus leave the field of view during rotations~(Figure~\ref{fig:mobile-device-issues}(b)). The authors suggest to alter the control-display ratio to avoid this limitation, but such an indirect mapping would become less natural than direct tangible manipulation.

One way to completely eliminate the separation, and its associated problems, would be to position the mobile device \emph{onto} the virtual object before starting the manipulation. However, a typical rendering process would clip half of the virtual object in this situation---as well as everything else in front of the screen---making it difficult to manipulate the object properly~(Figure~\ref{fig:mobile-device-issues}(c)). Additionally, most current mobile devices come in the form of flat one-sided displays. Even if clipping was not an issue, this form factor would still result in loss of visual feedback during manipulation. Therefore, we can identify two additional requirements: the portable and self-contained device should provide visual feedback on its entire surface, and should be able to display virtual objects without clipping.

\subsection{Fish-Tank~VR and geometric displays}

When rendering a 3D~scene, the display is generally considered as the viewpoint on the virtual world. This causes clipping whenever a virtual object crosses the screen plane. Fish-Tank Virtual Reality~(FTVR) provides a solution to address the clipping problem. FTVR turns the display into a ``window'' seen from the viewpoint of the user~\cite{ware93}. This allows virtual objects to appear behind, onto or in front of the display surface~\cite{francone11}. There are two main ways to achieve this effect: stereo rendering, head-coupled perspective, or a combination of both \cite{ware93}. According to Ware~et~al.~\cite{ware93}, head coupling is a much stronger clue than stereo rendering, making it the preferred solution to implement this technique.

As explained before, a flat one-sided display is not the best shape for use as a tangible handle. A number of works have proposed volumetric objects equipped with multiple displays, capable of providing visual feedback on their entire surface. Some of these objects take advantage of~FTVR to give the illusion of a 3D~space inside the device. They are called \textit{geometric displays}~\cite{stavness10}. One of the first examples is the MEDIA cube~\cite{inami97}: multiple LCD~displays were arranged in a box shape, and combined with head coupling to give the illusion that a virtual scene was inside the box. The CoCube~\cite{brown03} is a tangible cube that produced the same illusion when seen through a head-mounted display~(HMD). Unlike the previous example, this tangible cube was freely manipulable by the user. The Cubee~\cite{stavness06} is a cubic device that achieved the same goal without a~HMD, by using integrated displays and a head tracker. The pCubee~\cite{stavness10} is an evolution of Cubee that was made smaller and more portable. A similar device is gCubik~\cite{lopez-gulliver09}, which uses autostereoscopic rendering instead of head tracking. Most of the above devices have a cuboid shape, as this is the easiest way to arrange conventional rectangular displays, but other shapes are possible---such as arbitrary polyhedra~\cite{harish13} or even a sphere~\cite{ferreira14}.

Geometric displays appear to meet all our requirements: they can be made small, portable and self-contained, they provide visual feedback on all their surface, and with FTVR they can display virtual objects without clipping. However, previous work on this subject was mainly focused on the feasibility of such displays. Although significant technological contributions have been made, the potential of such devices for 3D~interaction remains comparatively little explored. For example, the pCubee device~\cite{stavness10} is one of the few works in which interaction techniques are mentioned---yet those are limited to visualizing 3D~objects and tumbling them inside the display, while object selection requires an external input device. In addition, there seems to be a lack of examples of positioning the geometric display itself within a larger virtual scene, which is an essential part of our concept. Even though the pCubee can be used to navigate in the virtual scene, this is accomplished through an indirect velocity-based mapping. In contrast, our concept is based on a full 1-1~mapping between the real world and the virtual world, which has greater potential for direct 3D~interaction. In this work, we offer a new perspective on geometric displays, by considering such a device as a portable and self-contained~TUI in its own right, capable of both displaying a virtual scene and serving as a tangible handle within the scene itself.

\section{Concept: a tangible volume}

\begin{figure}[t]
  \centering
  \includegraphics[width=.9\linewidth]{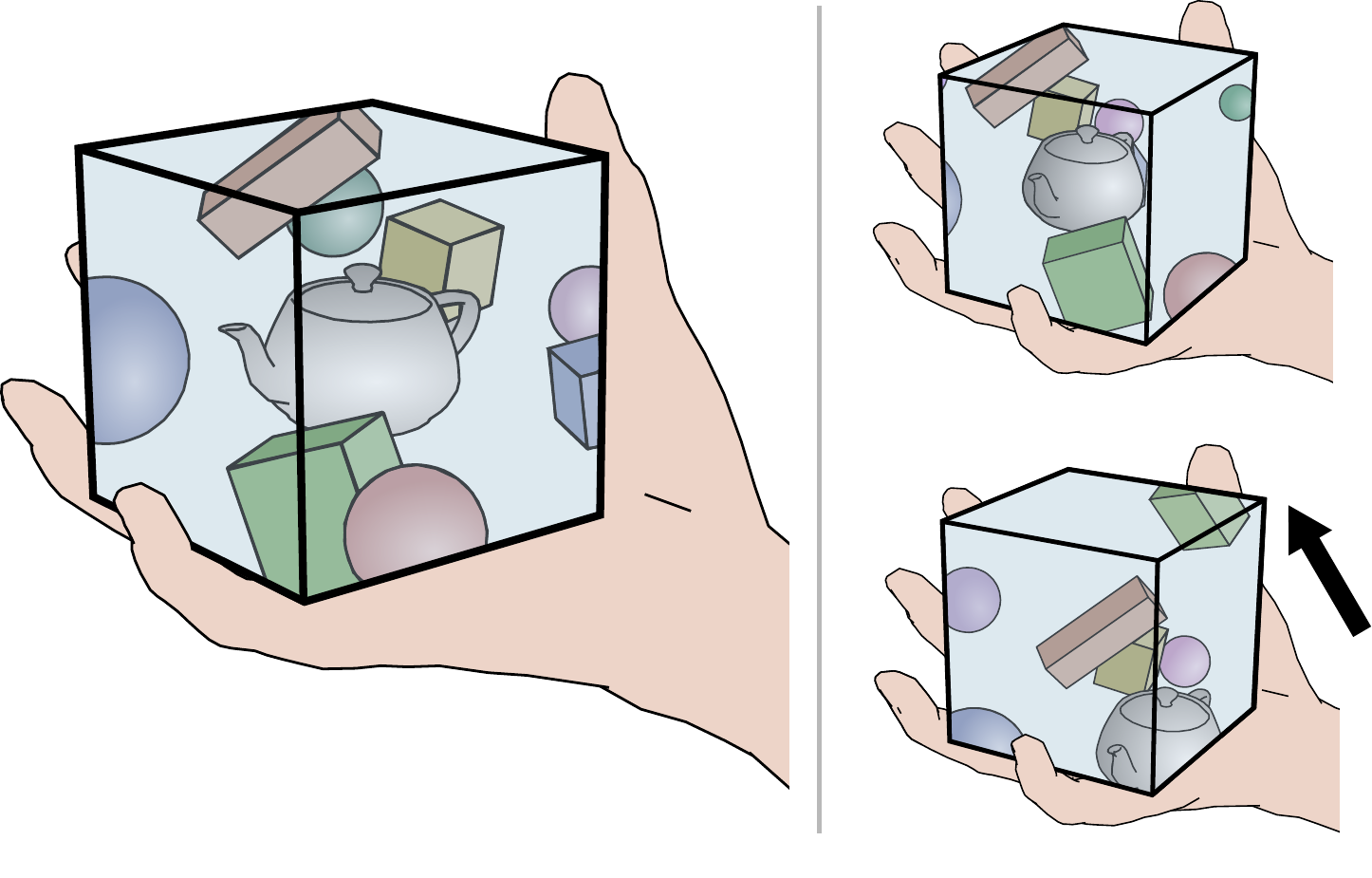}
  \caption{Illustration of our concept of a ``tangible volume''. Part of a larger virtual scene can be seen through the tangible object held by the user. On the top right, the tangible volume is observed from a different angle. On the bottom right, the tangible volume has been translated in space and now encloses a different part of the virtual~scene.}
  \label{fig:concept}
\end{figure}

Based on the above considerations, we introduce the concept of a ``tangible volume''. A tangible volume is a single physical object, sufficiently small and lightweight to be held in the hand. The surface of this object is entirely covered in screens on which the virtual scene is displayed. The perspective of each screen is adjusted to the user's head position. As a result, part of the virtual scene appears ``through'' the object. The object is also tracked relative to the real world. Users can reach other parts of the virtual scene by moving the object in real space~(Figure~\ref{fig:concept}). At any point, the physical boundaries of the object ``enclose'' a corresponding part of the virtual scene. Therefore, this object is a \emph{tangible} representation of a \emph{volume} of virtual space, that can be held in the hand and directly positioned into the virtual scene.

We use this tangible volume as an interaction device for \mbox{3D~object} manipulation. First, the tangible volume is positioned \emph{onto} a virtual object in the scene. From there, the virtual object is attached to the volume. The virtual object then follows the tangible volume in 3D~space, as if it was directly held in the hand~(Figure~\ref{fig:teaser}). Thus, there is no separation between the virtual object and the tangible~handle.

Our concept integrates all input and output into a single handheld object, used both to visualize the virtual scene and to manipulate virtual objects. Unlike the alternative approach of using a mobile device as a tangible handle, our interface also eliminates the separation from the manipulated virtual objects. Our concept thus constitutes a fully portable and self-contained interface for 3D~object manipulation which avoids the problems described before.

\subsection{Object selection by grasping}

Having an interface made of an unique tangible object has clear \mbox{benefits} for portability, but also raises the question of how to interact with more than one virtual object. Such an interface must provide a way to attach and detach virtual objects from the single tangible~handle.

In~TUIs made of multiple tangible objects, each tangible object can be linked to a different virtual object. This allows the user to interact easily with multiple virtual objects, simply by manipulating the corresponding tangible objects as desired. This is called ``space multiplexing''~\cite{fitzmaurice96}. In our case, there is only one tangible object available to manipulate an arbitrary number of virtual objects. Therefore, the user has to \emph{select} which virtual object should be linked to the tangible object at any given time. This is called ``time multiplexing''~\cite{fitzmaurice96}. Space multiplexing is considered more desirable than time multiplexing, due to the lack of an object selection step which improves efficiency and lowers cognitive load~\cite{fitzmaurice96}.

However, let us consider more closely what selection means in the specific case of tangible objects. Even though some interfaces allow the user to interact with multiple tangible objects concurrently, this is ultimately limited by the human capabilities. Tangible manipulation is typically performed with the hand(s). This means that all manipulation is accomplished through at most two effectors: the hands themselves. Any tangible object first has to be grasped with the hand in order to interact with the corresponding virtual object. Therefore, there is still an implicit selection in space-multiplexed~TUIs that occurs when an object is grasped with the hand.

From that perspective, one of the main advantages of space multiplexed~TUIs is not actually the lack of a selection step, but rather the fact this selection is implicit and does not require thinking. In other words, reaching for an object and grasping it with the hand constitutes a \emph{natural} way to select~it.

\begin{figure*}[t]
  \centering
  \includegraphics[width=\textwidth]{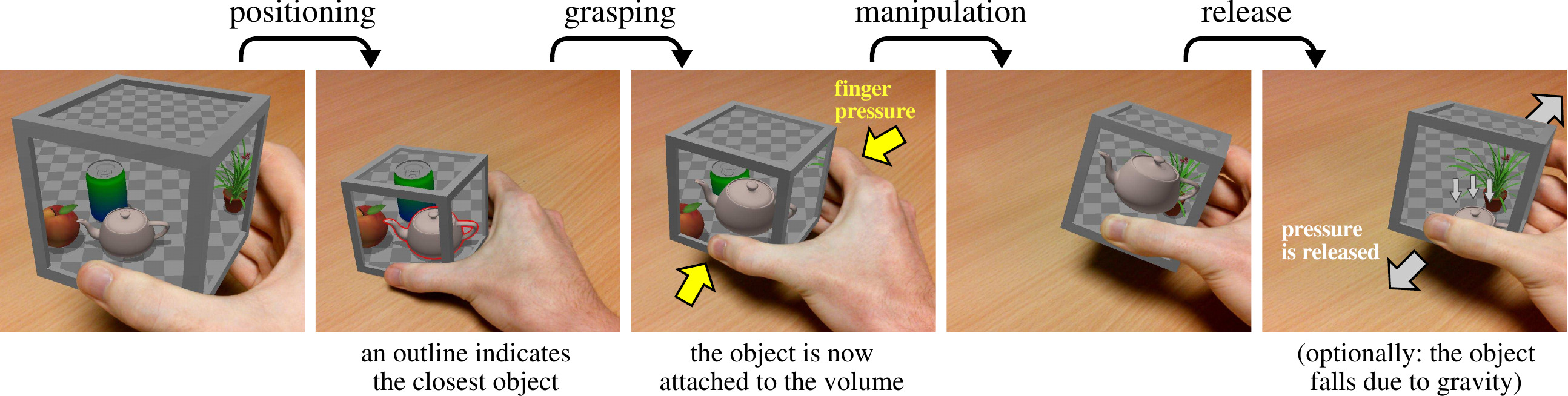}
  \caption{Illustration of the different steps of object manipulation in our interface concept. First, the tangible volume is positioned onto a virtual object. To disambiguate between nearby objects, an outline indicates which object is both inside the volume and closest to its center. This object can be grasped by pressing the fingers on the volume. The object is then attached to the volume, and can be directly moved alongside the volume in 3D~space. The object is detached when finger pressure is released. If virtual gravity is enabled, it then falls to the ground.}
  \label{fig:grasping}
\end{figure*}

\subsubsection*{Grasping metaphor}

The tangible volume provides a unique opportunity to reproduce this form of selection with virtual objects: by considering the volume as an extension of the hand. As shown above, the tangible volume can be moved in space and positioned around a virtual object. Since the volume is held in the hand, and is surrounding a virtual object, the hand is also surrounding the virtual object.

We thus designed a selection technique that consists in \emph{pressing the fingers} on the tangible volume to ``grasp'' the virtual object located inside~(Figure~\ref{fig:grasping}). Because the tangible volume is already held in the hand, the grasping technique is only triggered when finger pressure exceeds a given threshold. This threshold allows to differentiate a conscious grasping action from normal manipulation of the tangible volume itself. Similarly, releasing the virtual object is done by releasing finger pressure below this threshold, while keeping hold of the tangible volume.

With our grasping metaphor, picking up a virtual object becomes very similar to picking up a real-world object: placing the hand around an object, and pressing the fingers to grasp it.

\subsection{Disambiguating between virtual objects}

A challenge that arises from our grasping metaphor is that fingers cannot penetrate the volume. When virtual objects are close to each other, more than one may be located inside the volume bounds. Since fingers cannot be directly used to grasp the desired object, there must be a way to indicate which object will be selected among those located in the volume. One~possible solution is to display an outline around the object that is closest to the center of the volume~(Figure~\ref{fig:grasping}).

\subsection{Bimanual manipulation}

Space-multiplexed~TUIs allow the user to manipulate different virtual objects in each hand. Since our interface is made of a single object, this form of bimanual manipulation is not directly supported. One solution could be to use two tangible volumes---one for each hand. Since our grasping metaphor makes the tangible volume an extension of the hand, and all manipulation is accomplished through the hands, this solution would approximate space-multiplexed bimanual interaction. However, the interface would then again consist of multiple pieces, which we specifically wanted to avoid.

Still, as a single object, our interface supports another form of bimanual interaction: manipulating a single virtual object with both hands. This can be accomplished by holding the tangible volume with two hands, and pressing the respective fingers on opposite sides of the~volume.

\subsection{Simulated physics}

In his dissertation on Graspable User Interfaces, Fitzmaurice~\cite{fitzmaurice96}, citing~Norman~\cite{norman93}, argues that the natural laws of physics that affect tangible objects help the user during manipulation. Indeed, in graspable~UIs, releasing an object from the hand makes it drop to the floor due to gravity. This familiar behavior can help understand and predict its motion.

However, in our interface, the tangible volume always remains in the hand. Releasing a virtual object simply detaches the virtual object from the volume. Without further intervention, the virtual object would remain there---floating in space. We can obtain a more realistic behavior by adding simulated physics to the virtual scene. With physics, the released virtual object falls to the virtual ground when released~(Figure~\ref{fig:grasping}). Simulated physics also prevents virtual objects from moving through each other. This reinforces the illusion that they are solid, and thus can be grasped and manipulated directly.

It should be noted that virtual physics is not always desirable. For example, a complex manipulation task may have to be decomposed in several steps in mid-air (clutching). With virtual gravity, releasing the virtual object between each step would cause it to drop from its intermediate positions. Virtual physics enhances realism, but is not necessarily appropriate for every application.

\section{Implementation}

Having presented our concept as we envision it, we discuss here its technical feasibility given the current state of technology. We also describe a partial prototype, used in the two studies presented in the next sections.

\subsection{Technical feasibility}

Our concept has three main technological requirements. The first requirement is to cover the surface of a volumetric object with screens, as well as pressure sensors. Since the device is meant to be portable and self-contained, it should also contain enough processing capabilities to render a 3D~scene on multiple screens at interactive frame-rates, with enough battery power to operate for a sufficient duration. In view of the current advancement of geometric displays, as demonstrated by previous work, and the rapid pace of technological progress in the mobile device industry, this first aspect appears very close to be achievable.

The second requirement is to track the user's head position relative to the device, which is important to produce the fish-tank effect. In our concept, tracking requirements are made more challenging by the fact the device should be self-contained. Previous work on portable geometric displays (e.g.~\cite{stavness10}) relied on magnetic trackers, one attached to the user's head and the other to the object itself, along with an external tracking base. Assuming the base can be embedded in the object or in the head-worn device, and the head-worn device is comfortable and unobtrusive enough (e.g.~a pair of glasses), this solution represents a good compromise given the current state of technology. For a truly self-contained device, small cameras could be embedded on the surface of the object in order to optically track the user's head. Although current optical head tracking would not reliable enough in such a configuration, this solution remains theoretically achievable in the future.

The third requirement is to track the device relative to the real world. This is needed to produce the illusion that the virtual scene is fixed relative to the real world, which allows users to perform all the manipulation in real space. In a truly self-contained interface, there cannot be any external tracker to provide the real-world reference frame. Thus, the only solution is to use inside-out tracking from the device itself. Recently, the Tango project\footnote{\url{https://www.google.com/atap/project-tango/}} demonstrated how a device can track its own motion relative to the environment with an integrated infrared camera. Several of these cameras could be embedded on the surface of the device in order to provide continuous tracking of its real-world motion.

\subsection{Partial prototype}

\begin{figure}[t]
  \centering
  \includegraphics[width=\linewidth]{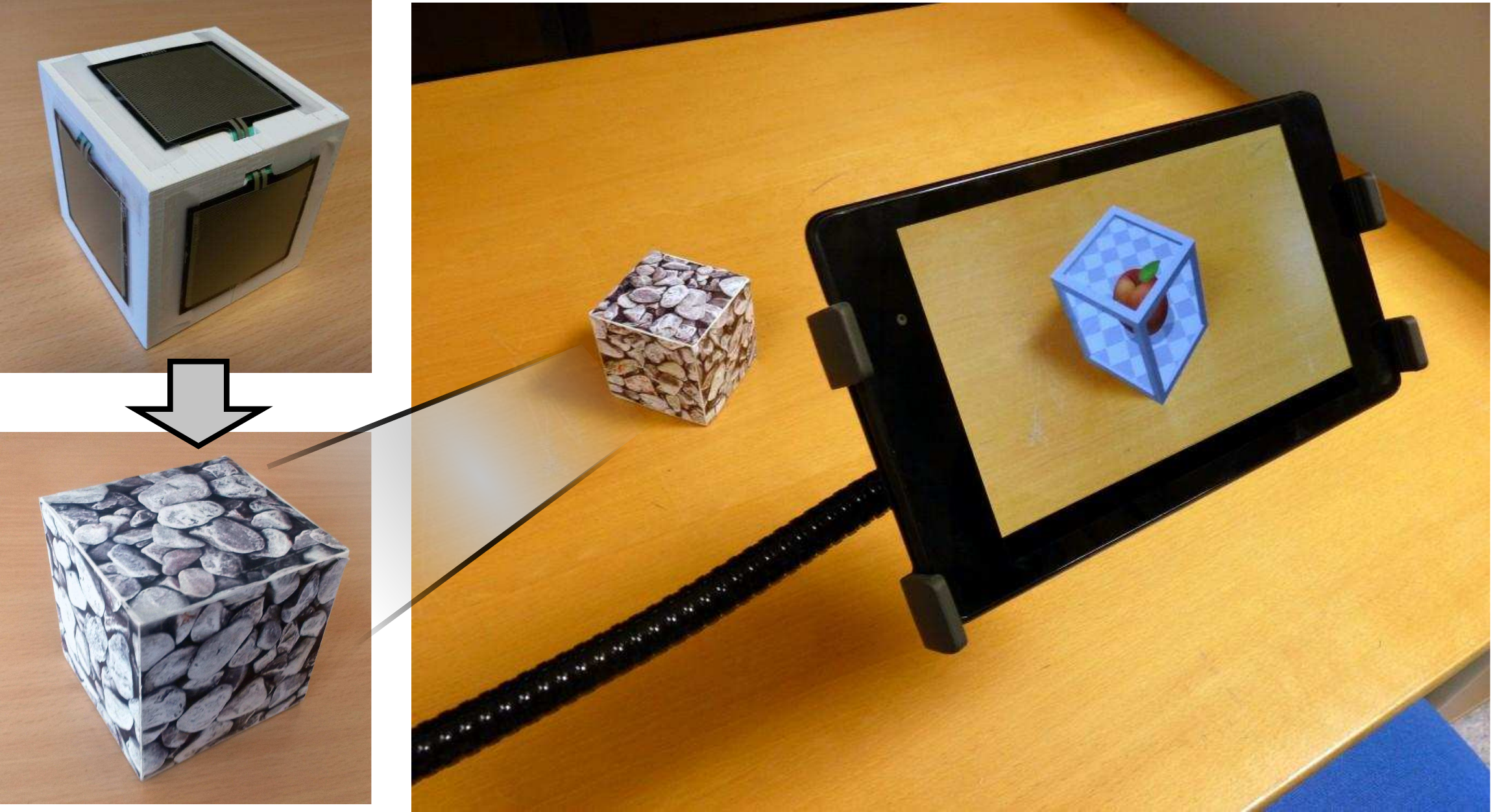}
  \caption{A partially simulated implementation of our concept, using augmented reality. A physical cube serves as the tangible volume. We covered its faces with six pressure sensors, then with AR~markers. A tablet, raised to eye level, turns the markers into virtual screens. Users manipulate the cube by looking through the tablet, as if they were directly holding the real device.}
  \label{fig:proto}
\end{figure}

In order to study several usability aspects of our concept, we designed a preliminary prototype. As said before, our work primarily focuses on the interaction capabilities afforded by a tangible volume, rather than on the hardware side. Therefore, we used augmented reality~(AR) to simulate some hardware aspects of our concept.

One may argue that existing geometric displays could have served as a basis to conduct user studies, without requiring AR~simulation. However, existing implementations described in previous work were all lacking in key aspects of our concept. Some of these devices are fixed installations, not meant to be moved by the user~\cite{inami97}. Others are somewhat movable, but are still tethered to a larger workstation~\mbox{\cite{lopez-gulliver09,stavness10}}. The tangible volume as intended in our concept is supposed to be fully portable and self-contained, and a tether wire could hinder manipulation~\cite{hinckley97}. Finally, some of these devices are only partially covered in screens~\cite{stavness10,stavness06}. Even though this is sufficient to demonstrate the technology, for complex object manipulation (especially rotations) it is important that all sides of the volume provide visual feedback. Simulating an interaction device to conduct user studies has been done in previous work~\cite{baricevic12}. This approach has the advantage of providing flawless rendering and tracking, which would be difficult to achieve in a research prototype but essential for the validity of user studies.

We chose a cubic shape for the tangible volume in our prototype. Even though other shapes could have been used, we chose a cube because it was easier to build while also being easy to manipulate by users. This cube was covered with AR~markers. We used a tactile tablet as an augmented reality window. When observed through the rear camera of the tablet, the faces of the cube were replaced with ``virtual screens'' that displayed part of the virtual scene~(Figure~\ref{fig:proto}). A frame was added around each virtual screen, to account for the fact that real screens would not be completely borderless. Using~AR provided implicit viewpoint tracking: since the cube faces were tracked by the camera, reversing this transformation produced an equivalent result. The cube was tracked relative to the real world by placing an additional AR~marker in the environment. We employed the Vuforia~framework\footnote{\url{http://www.vuforia.com/}} to track these objects. The tablet was attached to a raised stand, so that users could simply manipulate the object behind the tablet, as if they were directly looking at a cube equipped with screens.

\begin{figure}[t]
  \centering
  \includegraphics[width=.923\linewidth]{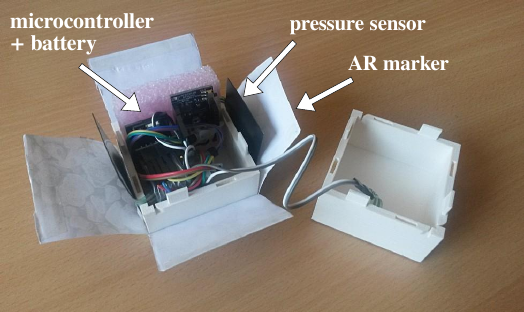}
  \caption{View of the electronic components inside the cube. The embedded microcontroller retrieves values from the pressure sensors on the cube surface, and sends them to the rendering software on the tablet through a wireless connection.}
  \label{fig:proto-inside}
\end{figure}

Our grasping technique was implemented in actual hardware, by attaching six flat pressure sensors (Interlink\textregistered{} FSR~406) to the faces of the tangible cube~(Figure~\ref{fig:proto}). The sensors were located under the~AR~markers, and thus were invisible to the users. They were driven by a microcontroller\footnote{\mbox{RFduino} RFD22102} embedded in the cube~(Figure~\ref{fig:proto-inside}). It was powered by a rechargeable battery, with a charging port hidden in one corner of the cube. The microcontroller continuously transmitted the pressure values to the rendering software on the tablet, through a wireless Bluetooth connection, at a frequency of~10\,Hz. Finally, we implemented physical simulation for the grasped objects with the~Bullet\footnote{\url{http://www.bulletphysics.org/}} physics engine.

In the rest of this paper, we use this prototype to investigate two important questions about our concept. First, we evaluate the intuitiveness of object manipulation with our grasping metaphor. In a second study, we evaluate the effects of the small field of view, which is limited to the surface of the tangible volume, on spatial~awareness.

\section{User study: object manipulation}

When interacting with the real world, grasping an object by pressing fingers on it, and moving it while maintaining finger pressure is a fairly natural procedure. However, it is unknown whether the exact same procedure remains natural when interacting with virtual objects through a tangible volume. We thus conducted an experiment to evaluate the \emph{intuitiveness} of object selection and manipulation in our~interface.

More specifically, we wanted to see if users can understand \emph{by themselves} with no prior explanations, how to grasp and move virtual objects with our interface. Of course, if some users do not succeed without help, this measure alone would not be sufficient to understand why. Therefore, we also wanted to determine at which step of manipulation those users would need guidance on their initial encounter with the interface.

\subsection{Participants and apparatus}

This study was conducted with unpaid 36~participants (from 20 to~52 years old, mean=29.5, sd=9.4). None of them had any prior knowledge of our interface concept or our grasping metaphor.

The apparatus was our AR-based prototype described in the previous sections. We slightly raised the pressure threshold to ensure that the grasping technique would only be triggered when truly intended by the participants.

\subsection{Procedure}

Before starting the experiment, we first told participants that it would consist in ``manipulating the cube''. We then explained why there was also a tablet in front of them: to ``simulate screens that should have been on the faces of the cube, but were not there due to technical limitations''. We followed by demonstrating how the cube turned into a cubic display when placed behind the tablet, and how a virtual scene could be seen through it. If our prototype had actual screens and head tracking, participants would likely have noticed all of this immediately. We did this short demonstration to ensure there was no confusion about our~AR simulation. However, we gave them \emph{no} explanation about how to interact with virtual objects.

\begin{figure}[t]
  \centering
  \includegraphics[width=0.9\linewidth]{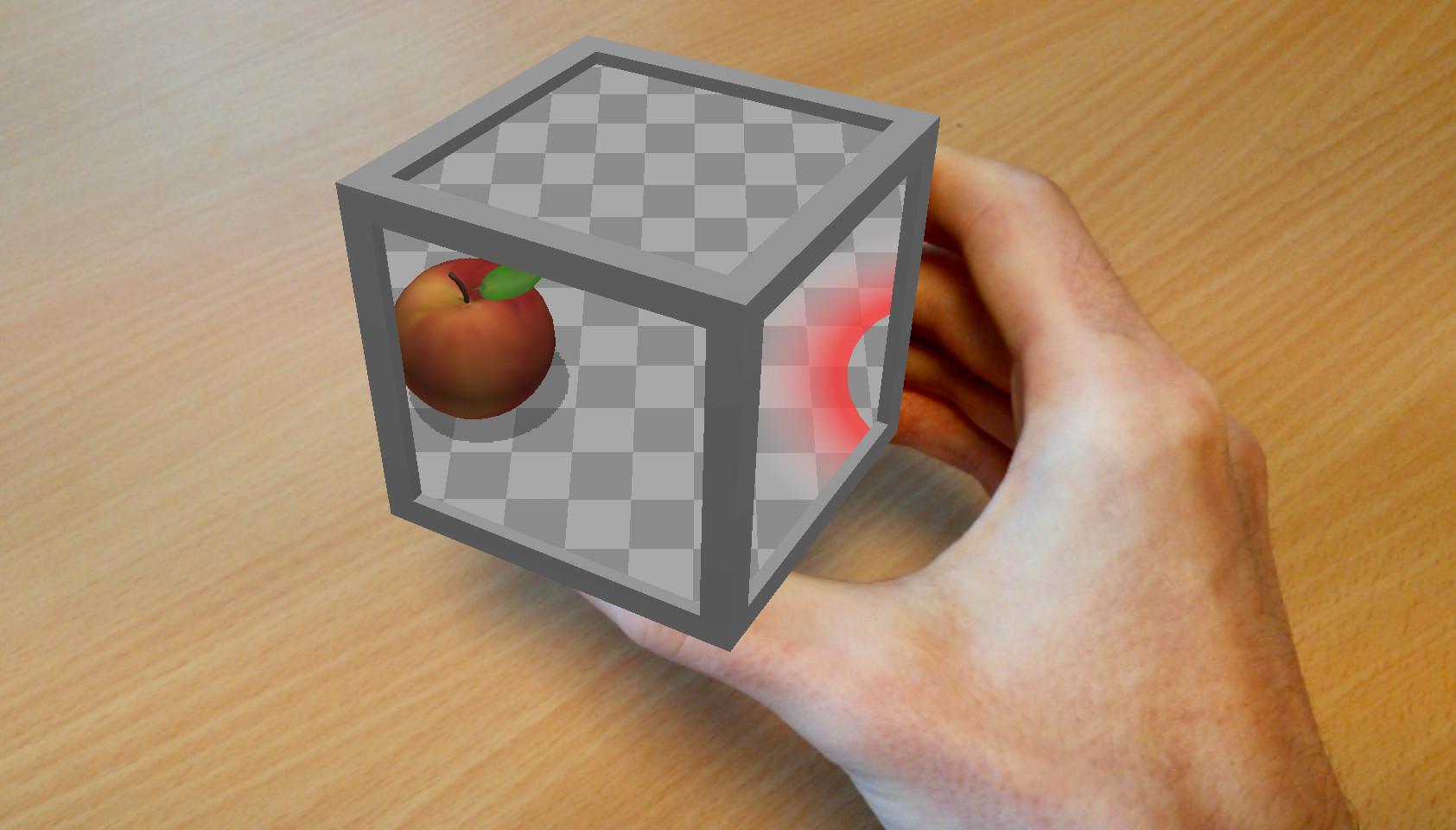}
  \caption{Screenshot of our first experiment, as seen by participants through the tablet. Note the lack of finger occlusion, due to the AR~simulation. The task was to pick up the virtual apple and move it to the red circle on the right.}
  \label{fig:results1-photo}
\end{figure}

We then introduced participants to the task: moving a virtual apple to a nearby target. The target was represented by a circle on the virtual floor~(Figure~\ref{fig:results1-photo}). Participants were told this task would have to be done ``with the cube''. They were also specifically told to ignore the tablet for manipulation. Finally, we asked them to discover how to do that ``by themselves, as far as possible''.

Obviously, we expected that some participants would not be able to complete the task without explanation. In order to understand how much help these participants needed, we designed a set of textual hints in increasing level of accuracy. During the task, there was a button that could be pressed to reveal a new hint. Each press on the button uncovered an additional hint on the tablet's screen, in the following order:
\begin{enumerate}[noitemsep]
\item ``Put the cube onto the apple''
\item ``Press the cube to grab the apple''
\item ``Move the cube while maintaining the pressure''
\end{enumerate}
These hints were specifically chosen to cover the different steps of object manipulation in our interface. Additionally, we chose to use textual hints rather than visual representations (e.g.~arrows) to avoid ambiguous interpretations. We explained the role of the hint button to participants, and strongly encouraged them to use as few hints as~possible.

\subsection{Results and discussion}

\begin{figure}[t]
  \centering
  \includegraphics[width=\linewidth]{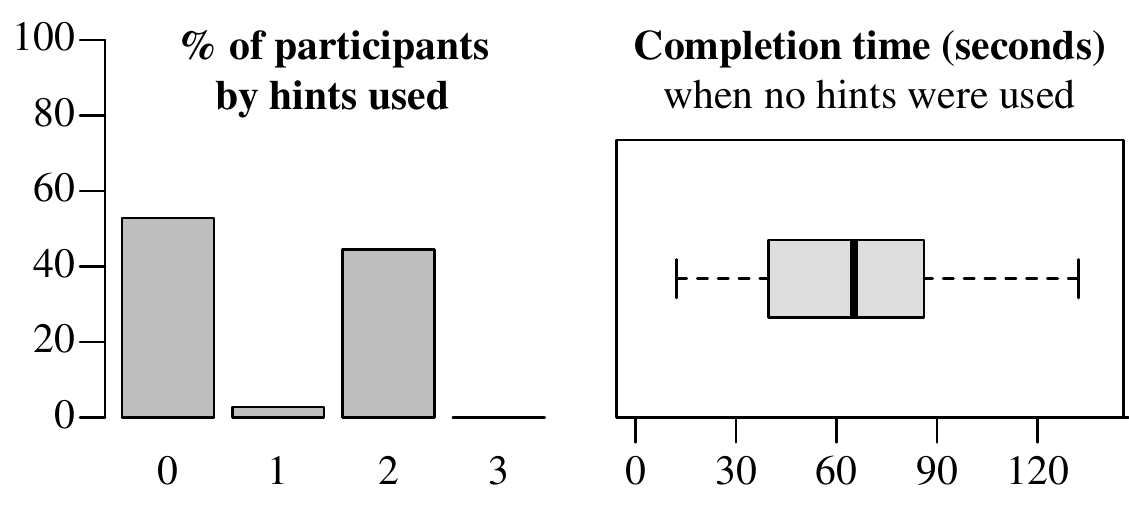}
  \caption{Number of hints and time needed by participants to discover how to manipulate a virtual object with our interface.}
  \label{fig:results1}
\end{figure}

Figure~\ref{fig:results1} shows the percentage of participants according to the number of hints they needed to complete the task. A total of~19 participants~(53\%) successfully completed the task without requesting any hint. Among the remaining participants, all but one~(45\%) succeeded with the first two hints. The third hint was never used. We also report the task completion times for the participants who accomplished the task without any hint. The mean completion time was~63.5\,s~(SD=34.3\,s).

More than half of participants discovered by themselves how to grasp and manipulate virtual objects with our tangible volume. This is an encouraging result, given that our interface is so much different from the way most people currently interact with virtual 3D~objects, and more generally interact with computers. At first glance, the completion times---about one minute on average---might seem long. However, this was the time needed to discover how to use a completely unfamiliar device \emph{and} accomplish a manipulation task for the first time. This, it seems that the idea of grasping a virtual object ``through'' a tangible volume was spontaneously considered by a majority of participants, and within a reasonable time.

For those who did not complete the task without help, the number of hints requested provides more insight into which parts of manipulation were troublesome. The first hint was designed to uncover potential difficulties in positioning the tangible volume onto the virtual object. Nearly all participants who requested hints were not helped with this first hint. Hence, positioning was likely not what prevented them from completing the task. Indeed, all participants quickly noticed that a outline appeared around the virtual object whenever it was inside the volume. Some of them, however, believed that the outline meant the object was already selected, and attempted to move it without pressing the fingers.

Among other attempted strategies, a surprising number of participants attempted to ``push'' the virtual object with the tangible volume. This may indicate they thought the sides of the volume would be solid in the virtual scene. Many tried to tap or flick their fingers onto the surface of the volume. In some cases, it was apparently an attempt to replicate the ``click'' metaphor, especially when they grasped the tangible volume as if it were a mouse and tried to click (and even double-click) on the top. In other cases, it was clearly an attempt to affect the virtual object through the volume, especially when flicking a finger against one of the side faces. Finally, some participants tried to fully enclose the cube in their hands. This may have been an attempt to grasp the virtual object. However, in doing so they lost visual feedback, and it was thus impossible to see if finger pressure was sufficient to trigger the grasping technique.

Nearly all participants who requested the first hint also requested the second hint to complete the task. The second hint was designed to uncover difficulties with the grasping metaphor itself. No participant ever asked for the third hint. For those who needed help, the grasping metaphor was therefore the main hurdle. The third hint was about moving the selected object by keeping it attached to the tangible volume. Since no participant requested this hint, none of them encountered any problem with this last step.

\section{User study: field of view}

Another challenge posed by our concept of a tangible volume is the limited field of view during manipulation. The tangible volume serves both as a display and as a tangible handle. Since our goal is to have a self-contained interface, the only display surface available is the surface of the volume itself. When the tangible volume is moved onto a virtual object to initiate manipulation, it also moves further away from the user's eyes, and its apparent size becomes smaller. Because of fish-tank rendering, the field of view on the virtual scene also becomes smaller. In contrast, moving the volume closer to the eyes would enlarge the field of view, but would also preclude direct~manipulation.

In a study on handheld~VR with mobile devices, Hwang~et~al.~\cite{hwang06} observed that the motion of the device in space was sufficient to compensate for a limited field of view. However, in our case the tangible volume is smaller than many typical mobile devices, and is held further away from the eyes during manipulation. Overall, the average field of view with our tangible volume is much smaller than in a typical mobile device configuration.

If this limited field of view had a significant impact on spatial awareness, this could be an disadvantage of our approach of having the display area restricted to the tangible volume itself. On the other hand, it is possible that the motion of the device in space and the fish-tank effect still compensate for this limitation in our case. We thus conducted a second user study to measure the impact of the limited field of view on spatial awareness.

\begin{figure}[t!]
  \centering
  \includegraphics[width=.82\linewidth]{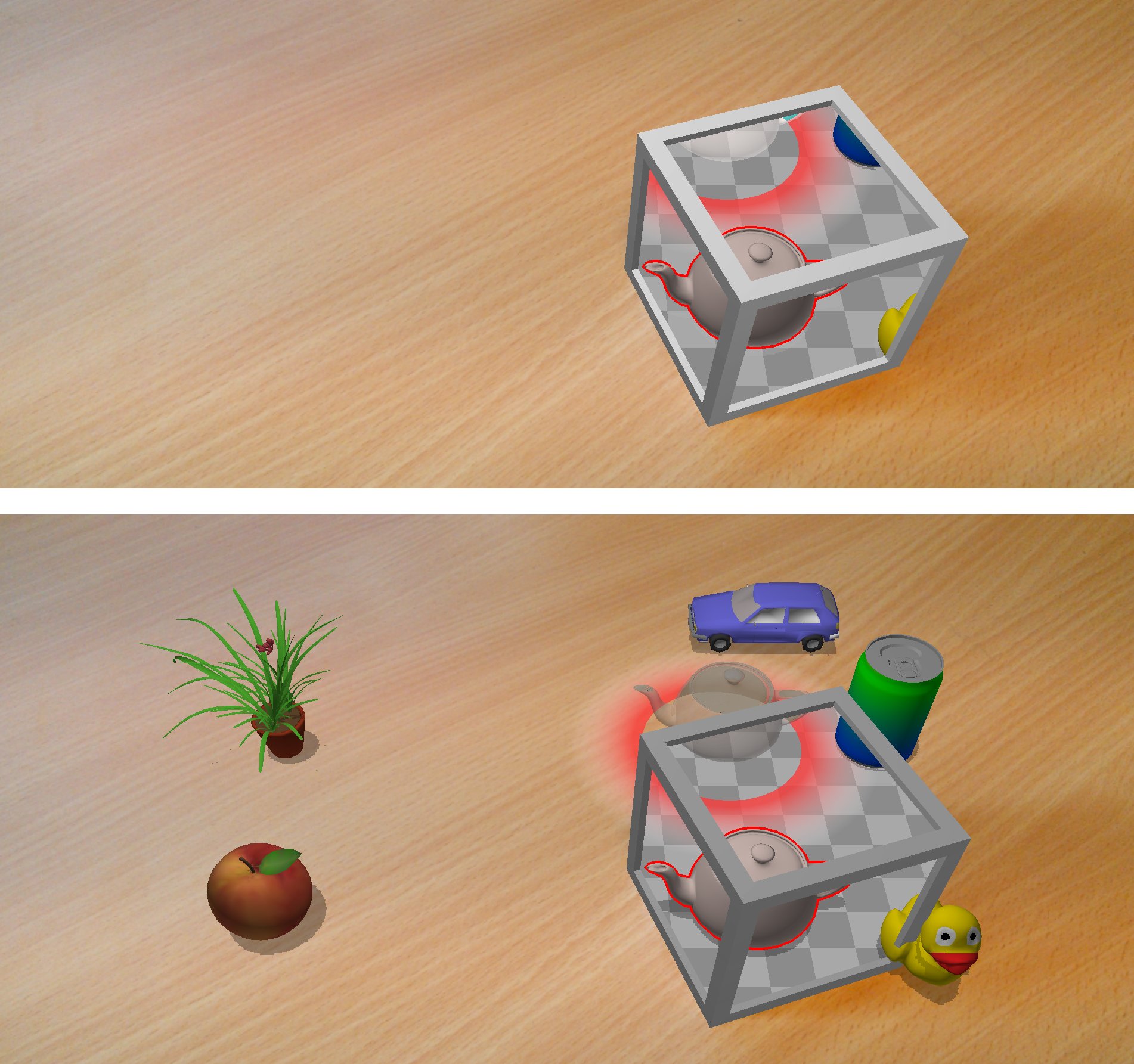}
  \caption{The two conditions in our second experiment. Top:~the normal condition, with the view limited to the tangible volume~(\emph{narrow~FoV}). Bottom:~the control condition, with the view extending outside the volume~(\emph{wide~FoV}).}
  \label{fig:results2-photo}
\end{figure}

\subsection{Participants and apparatus}

This study was conducted with 32~participants. All of them took part in our previous user study, and thus had prior experience with object manipulation using our interface. The apparatus was still our AR-based prototype, with the pressure threshold restored to normal.

\subsection{Procedure}

The task was to navigate a virtual scene consisting of 6~virtual objects, and to search for targets that would appear after a delay. Each successive target contained a silhouette of a given virtual object~(Figure~\ref{fig:results2-photo}), and this object had to be brought inside the target. After three of these manipulation tasks, the experiment stopped. Participants were then asked to indicate where they thought each virtual object was, by marking their location with small printed images. The goal was to evaluate how well they remembered the virtual scene, by comparing the reported object positions to the actual positions of virtual objects at the end of the experiment. The task itself was meant to simulate a typical use of the interface, by alternating navigation phases and manipulation phases.

The number of virtual objects in the scene~(6) was selected to be high enough for evaluating spatial memory, but not too large which would make the scene too complex. The objects themselves were chosen to be easy to recognize and remember. The objects started in seemingly random---but predefined---positions in the virtual scene. Three targets appeared during the experiment, for three different virtual objects. Thus, some objects were manipulated and some remained untouched, simulating a typical manipulation workflow. Targets appeared in seemingly random, but actually predefined locations. After~15\,s, a first target appeared. When the first manipulation was complete, a second target appeared after~10\,s. After the second manipulation step, a third target appeared after~12\,s. The third manipulation step was followed by a final delay of~20\,s. Then, the display went blank and participants were invited to report the last positions of virtual objects. To do so, they were given small pictures of the 6~virtual objects, and asked to position the pictures in front of them where they thought the virtual objects were located before.

In order to evaluate spatial awareness with the tangible volume, we compared it to an artificial situation in which virtual objects could be seen \emph{outside} the volume~(Figure~\ref{fig:results2-photo}). In this control condition, we simply modified the rendering pipeline in our AR-based prototype so that virtual objects were also rendered outside the tangible volume. Approximately half of participants~(17) performed the task in the \emph{narrow~FoV} condition, in which the field of view~(FoV) was limited to the tangible volume. The others~(15) performed the task in the \emph{wide~FoV} control condition, in which they could see the virtual objects outside the tangible volume during the entire~task. Two additional participants in the control condition were removed due to unusable~answers.

We used the \emph{Distance} metric proposed by Sharlin~et~al.~\cite{sharlin09} to quantify the difference between the positions reported by participants and the actual positions of virtual objects.

\begin{figure}[t]
  \centering
  \includegraphics[width=.95\linewidth]{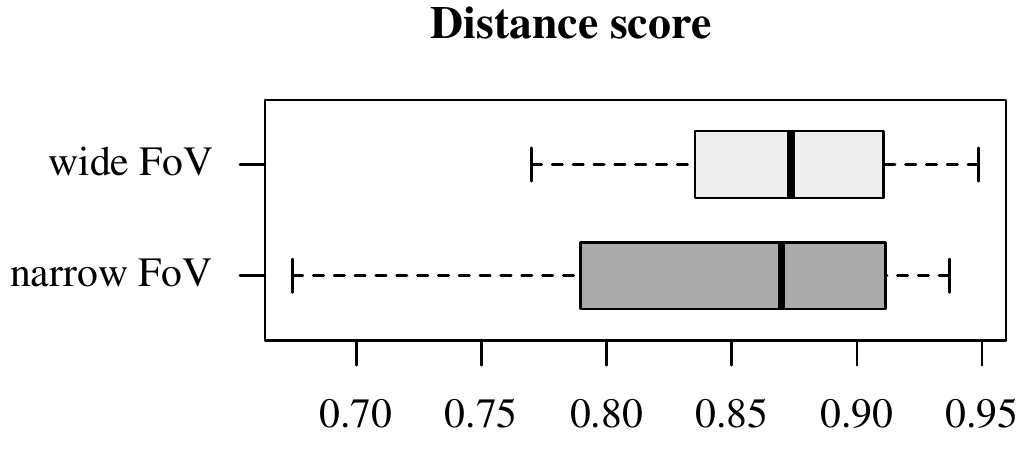}
  \caption{Correspondence between the positions reported from memory and the actual positions of virtual objects, based on the distance metric described by Sharlin~et~al.~\protect\cite{sharlin09}.}
  \label{fig:results2}
\end{figure}

\subsection{Results and discussion}

Figure~\ref{fig:results2} shows the distance scores computed from the participants' answers, grouped by condition. The mean score in the \emph{narrow~FoV} condition (view limited to the tangible volume) was~0.845~(SD=0.080). The mean score in the \emph{wide~FoV} control condition (view extending outside the tangible volume) was~0.869~(SD=0.057).

Although there were a few lower scores in the \emph{narrow~FoV} condition, the overall results did not seem to differ much. To confirm this, we used a two sample t-test to compare the means of each condition. The results (t(30)=0.99, p=0.32) yielded a difference of means in the range [-0.075,~0.026]~(95\%~CI). Thus, no practical difference between the scores of each condition was measured in our experiment.

From this result, it appears that the limited field of view in the \emph{narrow} condition was not an important disadvantage compared to the control condition. This suggests that having visual feedback limited to the surface of the tangible volume is not a disadvantage for manipulation. This is an important point for portability, because it eliminates the need for an external display surface.

This result also appears to confirm what was previously observed on mobile devices: moving a small display within the virtual scene may compensate for the limited field of view. Interestingly, in the narrow~FoV condition, we noticed that some participants did bring the tangible volume closer in order to obtain a wider field of view. In a fully implemented tangible volume, with actual screens and head tracking, this would be a natural action to do. However, we were surprised to see that the tablet, which acted as the user's eyes in our prototype, did not discourage some participants from using this strategy.

\section{Beyond manipulation: a tangible frontier}

We presented the advantages of turning a tangible object into a ``volume of space'' for creating a portable and self-contained~TUI. We also demonstrated its use for 3D~object manipulation, i.e.~translating and rotating virtual objects. But the concept of a tangible volume has more potential than just for manipulation.

The real world and the virtual world are inherently separated. No object or information can directly move between them. Interfaces are used to transfer information from one side to the other. For example, most interfaces are designed to transfer user input coming from the real world, and visual feedback coming from the virtual world. Thus, the physical medium of the interface acts as a \emph{frontier} that transfers information between the real and virtual worlds.

On a typical desktop computer, one such frontier is the monitor. The monitor constitutes a flat window that transfers visual information from the virtual world to the real world. However, user input comes from a completely separate location, usually the mouse and keyboard. With most current mobile devices, the frontier takes the form of a flat sheet of glass. This surface allows bidirectional transfer of information: light coming from virtual objects is transferred to the real world, and touch input on the surface is transferred to the virtual world. Since input and output happen on the same surface, the interaction is more direct. A mobile device is also portable: it can be freely moved in space to change the position of the frontier within both the real and virtual worlds. However, this frontier remains limited to a single planar surface, even though it is supposed to be an interface between two 3D~worlds.

In our concept, we extend this idea by giving the frontier a volumetric shape. The tangible volume represents a volumetric frontier between the real and virtual worlds. Its surface provides a physical medium on which sensors and actuators can be attached, in order to transfer information between the real and virtual worlds. Since this tangible surface can be freely moved within both the real and virtual space, \emph{every point of space becomes a potential interface} between the two worlds.

We demonstrated in the previous sections how this volumetric frontier can transfer information between the real and virtual worlds, in both directions. By attaching screens---a form of \emph{actuators}---to the tangible volume and using fish-tank rendering, light from the virtual scene appears to exit the virtual world, and continues its path into the real world toward the user's eyes. By attaching pressure \emph{sensors} to the tangible volume, force exerted by the user in the real world is directly transferred into the virtual scene.

\begin{figure}[t!]
  \centering
  \includegraphics[width=\linewidth]{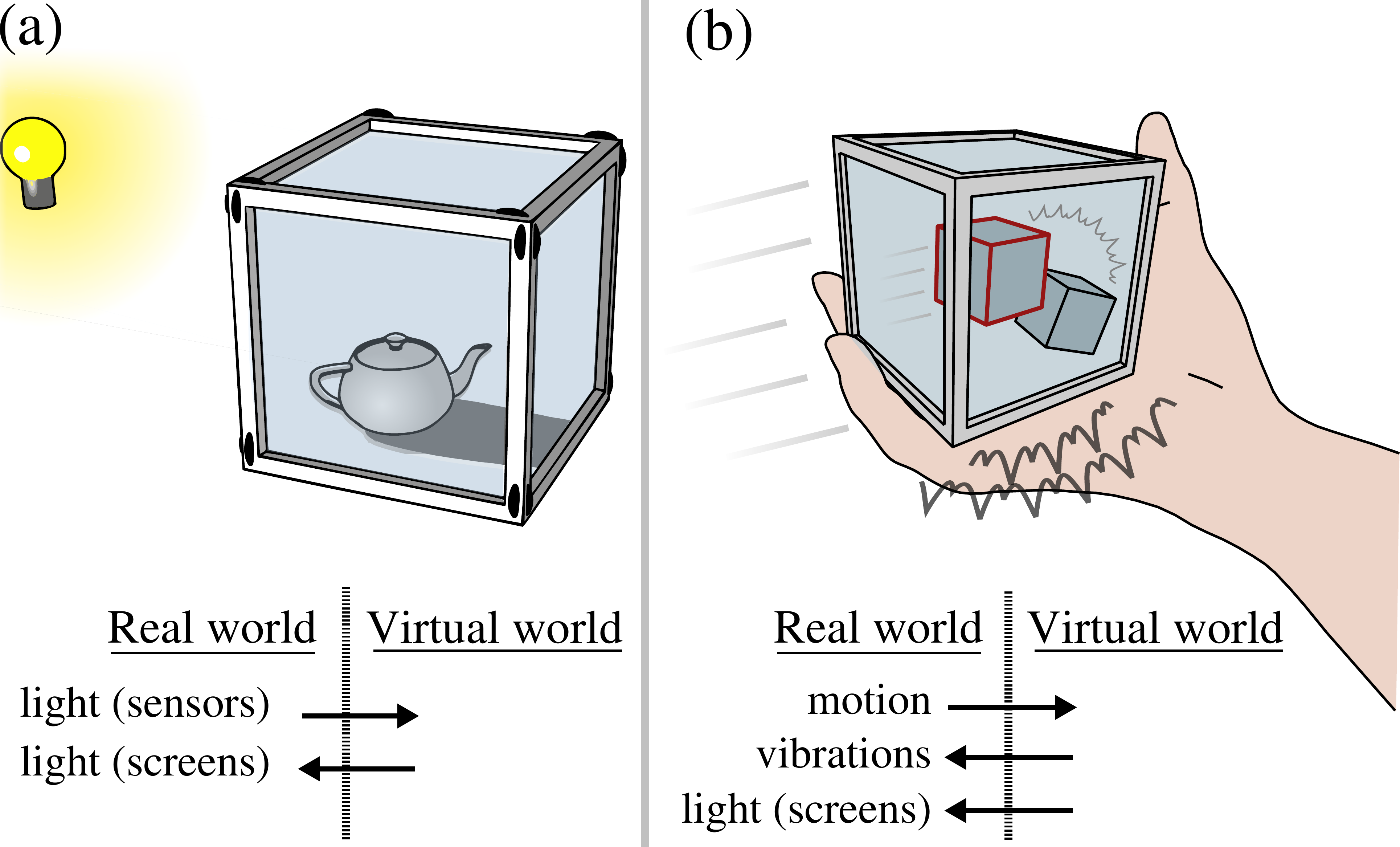}
  \caption{%
    The tangible volume can be seen as a handheld ``frontier'' that transfers information between the real and virtual
    words.\enskip%
    (a)\;by adding light sensors, light information can be transferred in both directions;\enskip%
    (b)\;by adding vibrators, haptic feedback (e.g.~collisions) from the currently selected virtual object can be transferred as well.%
  }
  \label{fig:frontier}
\end{figure}

Although we initially implemented these two modalities to support object manipulation, there is no reason why other modalities could not be transferred through the frontier as well. For example, light from the real world could be transferred to the virtual world, by attaching light sensors to the surface of the tangible volume. This would make it possible to render realistic lighting and shadows in the virtual scene~(Figure~\ref{fig:frontier}(a)). Similarly, sound can be transferred in both directions by integrating microphones and speakers into the tangible volume. Another sense that can be exploited is the haptic modality. Forces originating from the virtual object could be transferred to the real world. By adding vibrators to the tangible volume, collisions between virtual objects could produce haptic feedback directly on the user's hand~(Figure~\ref{fig:frontier}(b)). All of these forms of interaction can be implemented in a fully portable and self-contained interface, since the tangible volume provides the physical medium to attach any sensors and actuators. The shape of the frontier itself constitutes another possible interaction modality. Up to now, we only considered the case where the volume has a fixed shape (in our prototype: a cube). However, this shape could be made dynamic. By using flexible screens combined with underlying actuators, we can envision that this shape would change according to which virtual objects are located inside. This would bring the tangible frontier even closer to the virtual objects.

Examples where this concept of a tangible frontier could be used in practice include the interactive exploration of 3D~data, computer gaming, and virtual prototyping. For scientific visualization, the tangible volume can provide a direct access to the three-dimensional data for probing, selection, and data manipulation through the freely movable frontier. In gaming, innovative game concepts would become available by transferring multiple modalities (light, sound, haptics...) in both directions through the surface of the volume, increasing immersion and user engagement. Finally, the fact that tangible manipulation and other forms of 3D~interaction are offered by a single, portable and self-contained device can increase the opportunities for virtual prototyping. With the tangible volume alone, one could design a future product in virtual space and experience the sensation of holding it directly in the hand---with visual and haptic feedback. Furthermore, it would be possible to share the same experience with other designers or prospective customers by simply bringing along the tangible volume.

\section{Conclusion}

We introduced the concept of a ``tangible volume'': a fully portable and self-contained device for 3D~interaction, made of a single tangible object entirely covered with screens. This object represents a volume of the virtual scene, and can be positioned directly onto virtual objects. This makes 3D~interaction more direct than with previous approaches. We described an object manipulation technique that consists in grasping virtual objects ``through'' the volume and moving them in 3D~space. We created a partial prototype based on this concept, and used this prototype to investigate several usability questions. \mbox{Future work} could focus on three aspects. On the technical side, the first step would be to create a full implementation of our concept. This would require several improvements to the current technology. In particular, achieving reliable head tracking and environment tracking in a fully portable and self-contained device remains an important challenge. On the experimental side, a more complete implementation would allow to repeat user studies in conditions that more closely resemble a real device. For example, the influence of correct finger occlusion could be studied in a prototype with actual screens, as well as the effect of possible imperfections in tracking and rendering. Another question to investigate is whether adding stereo rendering could help the user position the volume within the virtual scene. Finally, our concept itself can be extended to other forms of interaction. We demonstrated how our tangible volume can be used for 3D~object manipulation. However, the concept of a tangible volume may be generalized as a volumetric ``frontier'' that can be positioned directly onto virtual objects. Future work could focus on exploring the full potential of this approach for other types of 3D~interaction with a virtual world.

%% if specified like this the section will be omitted in review mode
% \acknowledgements{The authors wish to thank \red{A, B, C}. This work
%   was supported \red{in part} by a grant from \red{XYZ}.}

\bibliographystyle{abbrv}
%%use following if all content of bibtex file should be shown
%\nocite{*}
\bibliography{article}
\end{document}